\documentclass[twocolumn]{aastex6}
\usepackage{natbib}
\usepackage{color}
\newcommand{\f}{\frac} 
\newcommand{\pare}[1]{\left(#1 \right)}
\fullcollaborationName{The Friends of AASTeX Collaboration}

\begin{document}


\title{Evolution of Cosmic Molecular Gas Mass Density from $z \sim 0$ to $z = 1-1.5$}


\author{Fumiya Maeda, Kouji Ohta, and Akifumi Seko}
\affil{Department of Astronomy, Kyoto University, Kitashirakawa-Oiwake-Cho,
 Sakyo-ku, Kyoto, 606-8502, Japan; fmaeda@kusastro.kyoto-u.ac.jp}







\begin{abstract}
We try to constrain the cosmic molecular gas mass density
at $z =1-1.5$ and that in the local universe by
combining stellar mass functions of star-forming galaxies and
their average molecular gas mass fractions against the stellar mass.
The average molecular gas mass fractions are taken from
recent CO observations of star-forming galaxies at the redshifts.
The cosmic molecular gas mass density is obtained to be
$\rho_{\rm H_2} = (6.8-8.8)~\times~10^7~M_\odot~{\rm Mpc}^{-3}$ at $z=1-1.5$ and
$6.7 \times 10^6~M_\odot~{\rm Mpc}^{-3}$ at $z \sim 0$
by integrating down to $0.03~M^\ast$.
Although the values have various uncertainties,
the cosmic molecular gas mass density at $z =1-1.5$ is about
ten times larger than that in the local universe.
The cosmic star formation rate density at $z \sim 1-2$ is also about
ten times larger than that in the local universe.
Our result suggests that the large cosmic molecular gas mass density at $z=1-1.5$
accounts for the large cosmic star formation rate density at $z \sim 1 -2$.
\end{abstract}




\section{Introduction} \label{sec:intro}
Cosmic star formation rate density (CSFRD) at $z\sim 1-2$ is considered to be about ten times larger than that in the present-day universe (e.g., \citealt{Lilly96}; \citealt{Madau}); i.e., galaxies were forming stars about ten times more active on average at the epoch.
What is the cause for this large CSFRD?
Since stars form in molecular clouds in galaxies, a simple explanation for this is due to a large cosmic molecular gas mass density (CMGD) at the redshift.
Another possible cause is a large star formation efficiency at the epoch.
Thus, revealing the CMGD is important to understand the evolution of CSFRD.
Based on semi-analytic model, \citet{Lagos11} and \citet{Popping14} calculated the cosmological evolution of the CMGD.
They showed the CMGD is 5 to 8 times larger at $z \sim 1 - 2$ than  at $z\sim 0$.
However, the molecular gas mass depends on prescriptions (such as pressure based, metallicity based) to evaluate the molecular gas mass fraction among various phases of the gas.
Hence, observational constraints are desirable.

In order to assess the CMGDs at the redshifts, a most straightforward way would be to derive the molecular gas mass function in the local universe and at $z \sim 1-2$, and to integrate them.
The molecular gas mass in a galaxy can be derived from its CO(1-0) luminosity.
For the local universe, the measurements of the CMGD using CO(1-0) luminosity were made (e.g., \citealt{Keres03,Obreschkow09a}).
Using an FIR- and a $B$-band selected sample of galaxies included in the Five College Radio Astronomy Observatory (FCRAO) Extragalactic CO survey, \citet{Keres03} derived a CO luminosity function and a CMGD by adopting a constant CO-to-${\rm H}_2$ conversion factor.
\citet{Obreschkow09a} applied variable CO-to-${\rm H}_2$ conversion factor (depending on CO luminosity and $B$-band luminosity) to the CO luminosity function by \citet{Keres03}.

For normal star-forming galaxies at $z \sim 1 - 2$,
CO observations of them have been very much time consuming and have been hard to be achieved.
Such a try was made by \citet{Walter14} as a blank sky survey in the Hubble Deep Field North using IRAM Plateau de Bure interferometer (PdBI);
they obtained CO luminosity functions at  $z$ = 0.34(CO(1-0)), 1.52(CO(2-1)) and 2.75(CO(3-2)) and derived the CMGD at each redshift.
Although they depicted the  cosmological evolution of the CMGD,
the uncertainty is very much large due to a small number of galaxies from which CO emission lines are detected.
More recently, \citet{Decarli16} conducted a spectroscopic survey with Atacama Large Millimeter/submillimeter Array (ALMA) in the Hubble Ultra Deep Field and the situation is improved.

As an alternative approach to the CMGD, \citet{Berta13} derived the molecular gas mass in main sequence galaxies at $z = 0.2 - 3.0$ by using a correlation of molecular gas mass and star formation rate.
But this approach does not fit to our motivation in this study.

The molecular gas mass in a star-forming galaxy can also be derived from the dust mass in the galaxy, by assuming a gas-to-dust ratio obtained in the local universe can be applied to high redshift galaxies.
However, a possibility that the gas-to-dust ratio at $z\sim 1-2$ is several times larger than those  in the local universe at the same gas metallicity is pointed out by \citet{Seko, Seko16b}, though \citet{Berta16} claim such evolution is  not seen.
In this study, considering the possibility of the evolution of gas-to-dust ratio, we derive the molecular gas mass without using the dust mass.

In this study, we take another approach to estimate a CMGD;
we combine the average gas mass fraction against stellar mass and the stellar mass function of star-forming galaxies.
Although the molecular gas mass fractions in CO detected star-forming galaxies have been individually derived,
the fractions tend to bias to more active star-forming galaxies with a high specific star formation rate
and gas mass fraction.
Thus, the unbiased average  gas mass fractions against stellar mass  of star-forming galaxies are needed.
Thanks to the recent large CO surveys,
the average  gas mass fraction against stellar mass of star-forming galaxies at $z \sim 0$ and $z\sim 1 - 2$ has been unveiling.
\citet{Saintonge11a} and \citet{Boselli} revealed the molecular gas masses of a few hundred of local star-forming galaxies.
\citet{Tacconi10,Tacconi} studied the molecular gas masses in main sequence star-forming galaxies at $z\sim 1-2.5$ with the IRAM PdBI and showed the average molecular gas mass fraction increases with decreasing stellar mass.
\citet{Seko} have studied the average molecular gas mass in main sequence galaxies at $z\sim 1.4$ with the ALMA and extended the trend to the lower stellar mass galaxies.

In this paper,
although the average molecular gas mass fraction against stellar mass
may still have an uncertainty,
nevertheless we try to constrain the CMGD at $z\sim1-1.5$ as well as that in the local universe
adopting this new approach mentioned above.
In section \ref{sec:fgas}, we describe the molecular gas
mass fraction and its dependence on the stellar mass, and
stellar mass function of star-forming galaxies.
Then in section \ref{sec:result}, we present the resulting CMGDs
obtained by combining these data, and
describe uncertainties on them.
In section \ref{sec:dis}, we compare the results with
the recent studies and model predictions,
and discuss implications.
In the same way as \citet{Madau}, 
initial mass function (IMF)  we use here is Salpeter IMF
 with an upper and lower mass
of $100~M_{\odot}$ and $0.1~M_{\odot}$, respectively.
Stellar mass, star formation rate (SFR) and molecular gas mass fraction 
[$f_{\rm mol} = M_{\rm mol}/M_{\rm star}$]
appear below are corrected to those with Salpeter IMF;
$M_{\rm Salpeter}$ (or SFR$_{\rm Salpeter}) = 1.7 \times M_{\rm
Chabrier}$ (or SFR$_{\rm Chabrier}$).
We adopt cosmology parameters of $H_0=70~{\rm km~s^{-1}~Mpc^{-1}}$,
$\Omega_{\rm M} = 0.3$, and $\Omega_{\Lambda} = 0.7$.

\section{Average Molecular gas mass fraction and stellar mass function} \label{sec:fgas}
\subsection{Average molecular gas mass fraction against stellar mass}
\citet{Tacconi10,Tacconi} used a sample of main sequence star-forming 
galaxies with $M_{\rm star}~\geq 4.3~\times~10^{10}~M_{\odot}$
and SFR $\geq 50~M_{\odot}~{\rm yr}^{-1}$ 
at $z =1-1.5$ (50 galaxies) observed with IRAM PdBI.
The molecular gas masses were derived by adopting 
the CO(3-2) (CO(2-1)) to CO(1-0)
luminosity ratio of 0.5 (0.6) and 
the CO-to-$\rm H_2$ conversion factor of
$4.36~M_{\odot}~\rm(K~km~s^{-1}~pc^{2})^{-1}$.
They also derived molecular gas mass fraction against 
the stellar mass 
for which CO emission is significantly detected.
Since the CO detected sample biases to higher specific SFR,
they derived the average molecular gas mass fraction 
in a stellar mass bin by correcting for the specific SFR of observed
galaxies.
Resulting gas mass fraction at $z= 1-1.5$ ranges from
$\sim 0.14$ at $M_{\rm star} \sim 3 \times 10^{11}~M_{\odot}$
to $\sim  0.34 $ at $M_{\rm star} \sim 6 \times 10^{10}~M_{\odot}$. 

\citet{Seko} used a sample of 18 randomly selected
main sequence  star-forming galaxies at $z \sim 1.4$.
Stellar mass and SFR covered are 
$M_{\rm star} \geq  4.0 \times 10^{9}~M_{\odot}$ and 
SFR $\geq 20~M_{\odot}~{\rm yr}^{-1}$, respectively,
expanding to the lower stellar mass. 
Further,  gas metallicity of the target galaxies was obtained
 with N2 method \citep{Yabe12,Yabe14}.
CO(5-4) observations were made with ALMA.
The molecular gas mass was derived  by adopting the CO(5-4)
to CO(1-0) luminosity ratio of 0.23 based on observations of main sequence star-forming galaxies (sBzK) of $z = 1.5$ by \citet{Daddi15}
 and the metallicity dependent CO-to-$\rm H_2$ conversion factor
(equation (7) by \citealt{Genzel12}).
To obtain the average value of the fraction at a stellar mass bin,
\citet{Seko} made stacking analysis including 
non-CO detected sample galaxies.
The fraction amounts to $\sim 0.72$ at the lowest stellar mass bin
of $2 \times 10^{10}~M_{\odot}$.

For the local star-forming galaxies, several surveys were conducted. 
CO Legacy Database for Galex Arecibo SDSS Survey 
(COLD GASS: \citealt{Saintonge11a}) measured the CO(1-0) luminosity
for a sample of 350 nearby  galaxies
($M_{\rm star} \gtrsim 1.7 \times 10^{10}~M_{\odot}$) using the IRAM 30-m
telescope.
CO(1-0) line was detected towards 222 galaxies, and 
they derived the molecular gas mass and its fraction 
against the stellar mass by adopting the  CO-to-$\rm H_2$ 
conversion factor of 
$3.2~M_{\odot}~\rm(K~km~s^{-1}~pc^{2})^{-1}$.
The CO detected galaxies are likely to 
consist mostly of late-type galaxies, by considering
their distributions in color, concentration, and stellar mass surface
density \citep{Saintonge11a};
they are considered to be  main sequence galaxies.
The average gas mass fraction among the galaxies
in a stellar mass bin ranges from
0.03 to 0.06; a slight tendency that the fraction increases
 with decreasing stellar mass is seen.

In $Herschel$ Reference Survey (HRS: \citealt{Boselli}), 
they extended to lower stellar mass 
($M_{\rm star} \gtrsim 1.7 \times 10^{9}~M_\odot)$). 
They also measured the CO(1-0) luminosity for 
a sample of 225 galaxies consisting of 57 E-S0a type galaxies and 
168 Sa-Im-BCD type galaxies. 
The detection rate is very low for early-type galaxies (16\%) 
and high for late-type galaxies (80\%).
They derived the molecular gas mass and its fraction against the
stellar mass by adopting the CO-to-$\rm H_2$ conversion
factor of $3.6~M_{\odot}~\rm(K~km~s^{-1}~pc^{2})^{-1}$.
The fraction is 0.19 at $M_{\rm star} \sim 2.4 \times 10^9~M_\odot$.
There is the same tendency as seen in \citet{Saintonge11a} 
that the fraction increases with decreasing stellar mass.

\begin{figure}[t!]
\includegraphics[width=80mm]{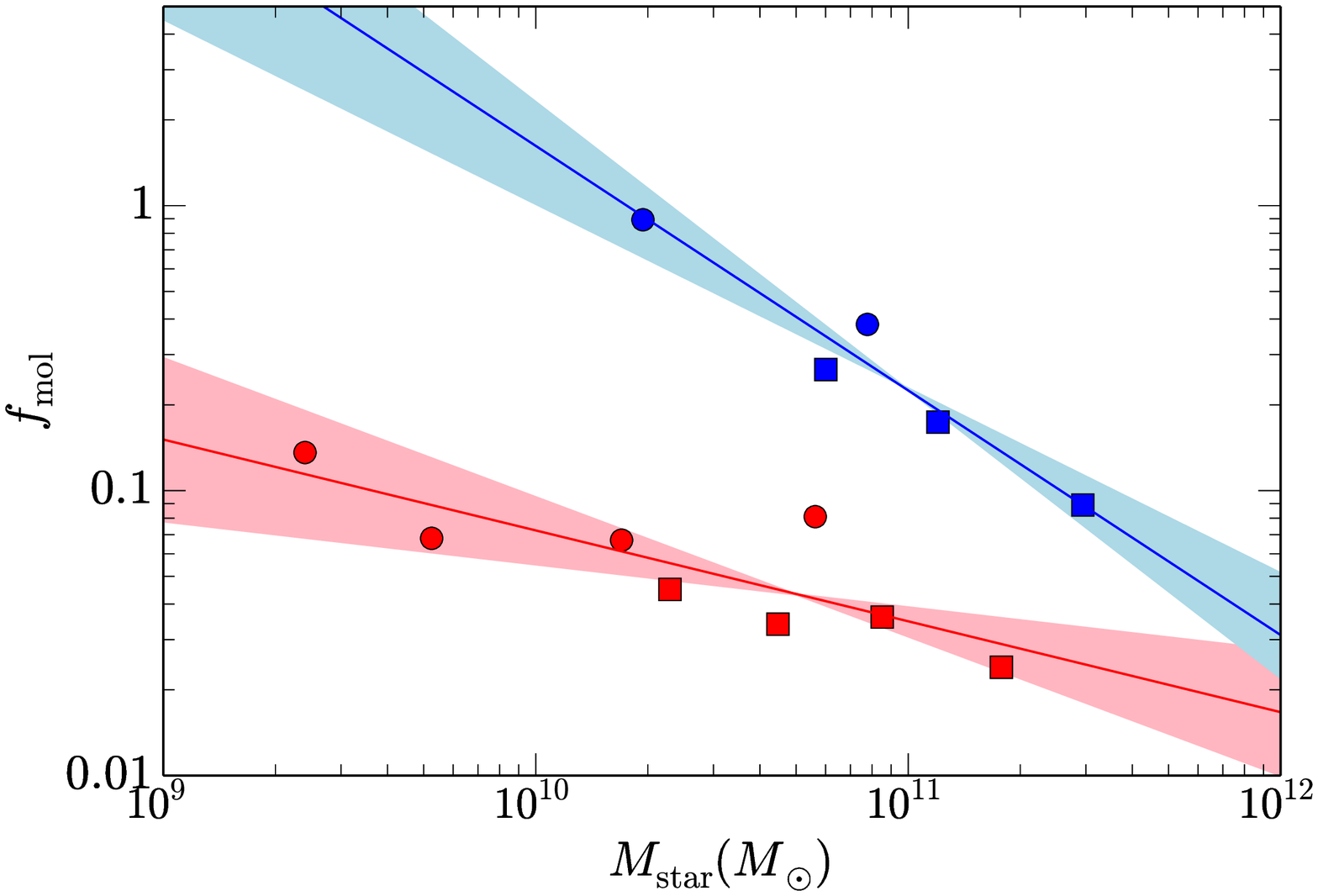}
\caption{\label{fig:fgas} Gas mass fraction ($f_{\rm mol} = M_{\rm mol}/M_{\rm star}$) against stellar mass.
For all sample data, we corrected to Salpeter IMF,
adopted the metallicity dependent CO-to-$\rm H_2$ conversion factor 
by \citet{Leroy11} and \citet{Genzel12} 
at $z \sim 0$ and $z =1-1.5$, respectively.
Filled blue squares refer to  the average values 
for the star-forming galaxies at $z = 1-1.5$ by \citet{Tacconi}.
Filled blue circles refer to the results of stacking analysis by \citet{Seko}. 
Filled red squares and circles refer to the average values in local star-forming galaxies 
by \citet{Saintonge11a} and \citet{Boselli}, respectively.
Solid lines represent the best-fit function by \citet{Popping} 
(equation (\ref{fmol})). 
Shaded regions show fitting uncertainty based on random realization
of the data points.}
\end{figure}

The gas mass fractions against the stellar mass are
summarized in Figure \ref{fig:fgas}.
Since different CO-to-H$_2$ conversion factors are used to 
derive the molecular gas mass in the data mentioned above,
we recalculated them by adopting the metallicity dependent 
CO-to-H$_2$ conversion factor by \citet[their Figure 6]{Leroy11} and
\citet{Genzel12} at $z \sim 0$ and $z =1-1.5$, respectively.
(Hereafter, we do not include the helium in the molecular gas mass.) 
To do this, we estimated the gas metallicity from the
mass-metallicity relation at $z\sim 0.1$ 
by \citet[their Figure 3]{Erb06} and 
at $z\sim 1.4$ by \citet{Yabe14} 
to use the same metallicity calibration.

Using these data, we fit the relation with the following function 
by \citet{Popping}: 
\begin{equation}
\label{fmol}
f_{\rm mol}(M_{\rm star}) =  \f{M_{\rm mol}}{M_{\rm star}}= \f{1}{\exp((\log M_{\rm star}-A)/B)} ,
\end{equation}
where $A$ and $B$ are constant parameters.
This function form was designed to match 
the molecular gas mass fractions against the stellar mass
at $0.5 < z< 1.7$ and local data including \citet{Saintonge11a}.
The best-fit values are $(A,~B)= (6.43,~1.36)$ and $(A,~B) = (10.24,~0.51)$ 
at $z \sim 0$ and $z =1-1.5$, respectively. 
Note that the original function form is for the fitting to the gas mass fraction of
$M_{\rm mol} / (M_{\rm mol} + M_{\rm star})$ and does not exceed 1.0. 
Uncertainties (1$\sigma$) of the best-fit functions are shown as shaded region in Figure \ref{fig:fgas}.
We derived the uncertainties by random realizations of the data points according to their errors.

\defcitealias{Tomczak}{T14}
\defcitealias{Mortlock}{M15}

\subsection{Stellar mass function of star-forming galaxies}
We use the stellar mass functions (SMFs) at $z = 1 - 1.5$
 by \citet[\citetalias{Tomczak}]{Tomczak} 
and \citet[\citetalias{Mortlock}]{Mortlock}.
At $z \sim 0$, we adopt the SMF of star-forming galaxies by \citet{Moustakas}.
The SMFs are also corrected for the IMF difference.

\citetalias{Tomczak} derived  galaxy SMFs over a redshift range 
of $0.2 < z < 3$ using $\sim$ 13,000 galaxies from 
the FourStar Galaxy Evolution Survey obtained in 
the Chandra Deep Field South, the Cosmic Evolution Survey,
and the Hubble Ultra Deep Field. 
\citetalias{Tomczak} derived the SMF down to about 1 dex lower 
stellar mass than those of the previous studies at $0.2 < z < 3$.
\citetalias{Tomczak} reached stellar mass of $6.3 \times 10^8~M_\odot$ at $z = 1.0 -1.5$.
They separated the full galaxy sample into star-forming and 
quiescent populations based on a rest-frame $U-V$ vs. $V-J$ diagram, 
and then derived SMF for respective populations.
They fitted the SMFs with double-Schechter function as
\begin{eqnarray}
\label{SMF}
\lefteqn{\Phi(M) dM = } \nonumber \\
& & e^{-M/M^\ast} \pare{\Phi_1^\ast \pare{\f{M}{M^\ast}}^{\alpha_1}+\Phi_2^\ast \pare{\f{M}{M^\ast}}^{\alpha_2}} \f{dM}{M^\ast}
\end{eqnarray}
where $M^\ast$ is the characteristic stellar mass.
We derived a SMF of star-forming galaxies at $z=1-1.5$ 
by combining the  best-fit double-Schechter SMFs
at $z=1.0-1.25$ and $1.25-1.5$ (their Table 2). 

\citetalias{Mortlock} derived SMFs over the similar redshift range
($0.3 < z < 3.0$) using a combination of 
the UK Infrared
Telescope Infrared Deep Sky Survey (UKIDSS)
Ultra Deep Survey (UDS), Cosmic Assembly Near-infrared
Deep Extragalactic Legacy Survey (CANDELS) UDS and CANDELS the Great Observatories
Origins Deep Survey-South survey data sets.
\citetalias{Mortlock} reached stellar mass of $3.2 \times 10^8~M_\odot$ at $z = 1 - 1.5$.
They selected SF galaxies based on $UVJ$ classification
(contamination by SF galaxies in quiescent population is estimated to the on average $\sim 2$~\%
by using 24~$\mu$m data), and derived SMF of star-forming galaxies
at $z = 1.0-1.5$ and fitted with the single-Schecheter 
function.
The uncertainties on the SMFs of \citetalias{Tomczak} and \citetalias{Mortlock} 
include the uncertainties of
the $UVJ$ classification
(i.e., contamination due to the color classification), the uncertainties in the SED modeling, the Poisson uncertainties, and
cosmic variance.

\citet{Moustakas} derived a SMF of nearby galaxies ($z=0.01 - 0.2$) 
using $\sim$ 170,000 galaxies from the 
Sloan Digital Sky Survey
(SDSS).
The SMF in the local universe reached stellar mass of $1.7 \times 10^9~M_\odot$.
They separated the galaxies into star-forming and 
quiescent populations based on whether they lie on or below 
the main sequence in SFR vs. stellar mass diagram.
We fitted the single-Schechter function and
obtained the best-fit parameters :
 $(\log M^\ast/M_\odot,~\alpha,~\log\Phi^\ast) = \ \
(11.07~\pm~0.02, \ -1.30~\pm~0.03~,~-2.98~\pm~0.04)$.

\section{cosmic molecular gas mass density} \label{sec:result}
Combining the dependence of the molecular gas mass fraction
on the stellar mass and the SMF, we derived the CMGD as  
\begin{equation}
\label{MGD}
\rho_{\rm mol} = \int_{M_{\rm min}}^{M_{\rm max}} f_{\rm mol } M_{\rm star} \Phi(M_{\rm star})dM_{\rm star},
\end{equation}
where $\Phi(M_{\rm star})$ is the SMF of
star-forming galaxies and $f_{\rm mol} $ refers to $M_{\rm mol}/M_{\rm star}$ (equation (\ref{fmol})).

Since we intend to compare the cosmic evolution of the
CMGD 
with that of the CSFRD,
obtaining the molecular gas masses
in the galaxies in the same stellar mass range as that for
the CSFRD is desirable.
\citet{Madau} derived  the CSFRD
by integrating luminosity function from $0.03~L^\ast$
where $L^\ast$ is the characteristic luminosity of the Schechter function.
This does not necessarily correspond to stellar mass exactly.
But considering the correlation between SFR and stellar mass,
i.e., main sequence for star-forming galaxies, it would be
reasonable to choose $M_{\rm min}$ as $0.03~M^{\ast}$ at each redshift.
From the best-fit $M^\ast$,
$M_{\rm min}$ in the local universe is $3.6~\times~10^9~M_\odot$
and that at $z=1-1.5$ is  
$1.7 \times 10^9~M_\odot$(\citetalias{Tomczak}) and
$2.2 \times 10^9~M_\odot$(\citetalias{Mortlock}).
As for  $M_{\rm max}$, we take $10^{12} M_{\odot}$.

\begin{figure}[t!]
\includegraphics[width=90mm]{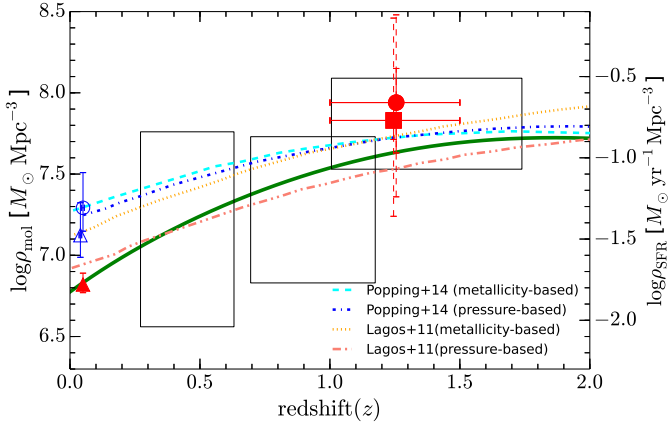}
\caption{\label{fig:cmgd} CMGD (left ordinate)
and CSFRD (right ordinate) against redshift.
Filled red circle and square refer to the CMGD
at $z \sim 1-1.5$ by adopting SMFs
by \citetalias{Tomczak} and \citetalias{Mortlock}, respectively.
Filled red triangle refer to the CMGD at $z \sim 0$.
The vertical solid error bars only include fitting uncertainties of the gas mass fraction and SMF.
Dashed error bars at $z \sim 1 -1.5$ show the possible systematic uncertainty of the CO luminosity ratio 
of a factor of 3 (see text).
These symbols are slightly shifted in the horizontal direction for clarity.
We adopt the metallicity dependent CO-to-$\rm H_2$ conversion factor.
Box frames show the results by \citet{Decarli16} 
based on the CO luminosity function.
They adopted the conversion factor of
$3.6~M_{\odot}~\rm~(K~km~s^{-1}~pc^{2})^{-1}$.
Open blue circle and triangle show the result by \citet{Keres03} and \citet{Obreschkow09a}, respectively.
\citet{Keres03} adopted the conversion factor of $4.8~M_{\odot}~\rm~(K~km~s^{-1}~pc^{2})^{-1}$.
\citet{Obreschkow09a} used the CO luminosity function by \citet{Keres03} but adopting the CO luminosity and $B$-band luminosity dependent conversion factor.
Semi-analytic model predictions for the CMGD
by \citet{Popping14} and \citet{Lagos11} are also shown.
Green solid curve represents the best-fit CSFRD by \citet{Madau}.}
\end{figure}

The resulting CMGD at 
$z \sim 0$ is $6.7_{-0.8}^{+0.9} \times 10^6 ~M_\odot~{\rm Mpc}^{-3} $ 
and that at $z = 1 - 1.5$ is
$8.8_{-4.4}^{+5.2} \times 10^7~M_\odot~{\rm Mpc}^{-3}$ (\citetalias{Tomczak}) and 
$6.8_{-1.3}^{+2.2} \times 10^7~M_\odot~{\rm Mpc}^{-3}$ (\citetalias{Mortlock}).
The results are shown in  Figure \ref{fig:cmgd}.
The results by adopting \citetalias{Tomczak} and \citetalias{Mortlock}
agree with each other within the error.
Here the uncertainties ($1 \sigma$) shown with solid error bars to the obtained values are calculated
from the uncertainties on the molecular gas mass fraction (equation (\ref{fmol})) and the SMF;
we ran $10^4$ realizations assuming the error distribution for the
best-fit values is Gaussian.

The CMGD at $z = 1-1.5$ is about ten times
larger than that in the local universe.
This seems to imply that the large CFRDS is due to the large
CMGD at $z~=~1~ -1.5$.
There are, however, many uncertainties in deriving the CMGD
other than the fitting error
on the molecular gas mass fraction and the SMF:
(1) CO luminosity ratio,
(2) CO-to-$\rm H_2$ conversion factor,
(3) $f_{\rm mol}$ in low stellar mass range,
(4) slope of main sequence, and
(5) contribution from Ultra-Luminous InfraRed Galaxies (ULIRGs). 
We discuss these uncertainties.

(1) To obtain CO(1-0) luminosity, \citet{Seko} assumed the 
ratio of 0.23 for CO(5-4) luminosity by \citet{Daddi15}.
\citet{Daddi15} reported that 
CO(5-4) emissions from three main sequence galaxies (sBzK) at $z = 1.5$
are more excited than that of our Galaxy,
but are not so excited as compared with M82.
Meanwhile,
\citet{Decarli16b} recently found the $J$-ladders of $z=1-1.6$ star-forming
galaxies are more similar to that of our Galaxy.

If the CO $J$-ladder is similar to that of M82 (e.g., \citealt{Carilli13}), 
the CO(1-0) luminosity would be lower by a factor of $\sim 3$.
If the ladder resembles to that of our Galaxy,
it increases by a factor of $\sim 3$.
Dashed error bars in Figure \ref{fig:cmgd} show the uncertainty.
For the CO(3-2) and CO(2-1) lines, the uncertainties are smaller than this.

(2) If we adopt a metallicity independent CO-to-$\rm H_2$ conversion factor of 
$3.6~M_{\odot}~\rm(K~km~s^{-1}~pc^{2})^{-1}$ (not including helium mass),
the CMGD decreases
23 \% (\citetalias{Tomczak}) and 25 \% (\citetalias{Mortlock}) at $z = 1 -1.5$
and 6 \% at $z \sim 0$.

(3) We extrapolated the molecular gas mass fraction to $M_{\rm star} =
1.7 \times 10^9~M_{\odot}$ at $z=1-1.5$ to cover the similar
stellar mass range used for the derivation of the CSFRD.
\citet{Saintonge13} and \citet{Dessauges15}
show the molecular gas mass fractions in lensed star-forming galaxies with
stellar mass of $2.5 \times 10^9~M_\odot \leq M_{\rm star} \leq 1.0 \times 10^{10}~M_\odot$ at $z = 1.5-3$ are
$f_{\rm mol} \sim 1-2$, supporting our extrapolation.
However, the average molecular gas mass fraction
in star-forming galaxies with the low stellar mass range
may be smaller,
because the low metallicity and the low gas column density
are likely to suppress formation of molecular component.
Although the CO-to-${\rm H}_2$ conversion factors we used are metallicity dependent, 
we also calculated the CMGD
assuming $f_{\rm mol} = 1.0~(f_{\rm mol} = 0.0)$ 
in $1.7 \times 10^9~M_\odot< M_{\rm star} < 2.0 \times 10^{10} M_\odot$.
Resulting CMGD is reduced by 25\% (44\%).
Further studies of the gas mass fraction 
in the low stellar mass range are desirable.

(4) If we consider that the slope of the main sequence is slightly flatter (${\rm SFR} \propto M_{\rm star}^{0.7}$),
$M_{\rm min}$ would be $0.007~M^\ast$. In this case,
the resultant CMGD increases $70 - 80$\% at $z = 1- 1.5$ and 18~\% at $z \sim 0$,
though the uncertainties of the fractions and SMFs are larger than those for the case of $0.03~M^\ast$. 

(5) SMFs of star-forming galaxies used here do not exclude ULIRGs,
but here we discuss the contribution to the CMGD from the ULIRGs.
The gas mass fractions of ULIRGs at $z = 1.4 - 1.7$ by \citet{Silverman15} are the same order as those of main sequence galaxies.
This can also be seen by comparing the stellar mass vs. SFR diagram (e.g., \citealt[their Figure 1]{Rodighiero}) and the molecular gas mass vs. SFR diagram (e.g., \citealt[their Figure 2]{Sargent}).
Considering the stellar mass of ULIRGs is about $10^{10} - 10^{11}~M_\odot$,
the number density of ULIRGs \citep{Goto} is more than ten times smaller than that of main sequence galaxies
with $M_{\rm star} = 10^{10} - 10^{11}~M_\odot$ ((\citetalias{Tomczak}) and (\citetalias{Mortlock})).
Hence, the contribution to the CMGD from ULIRGs is expected to be small.

The largest uncertainty seems to be the CO luminosity ratio,
and the observations in lower CO transitions are more desirable.

\section{Discussion and summary} \label{sec:dis}
The CMGD at $z = 1 -1.5$ obtained here is consistent with that
recently derived value based on the CO luminosity function at $z =1-1.7$
\citep{Decarli16}.
They derived CO(2-1) luminosity function 
and used the CO emission line $J$-ladder by \citet{Daddi15}.
They also conclude the CMGD at  $z =1-1.7$ is
$3 - 10$ times larger than that at the present-day universe.
The CMGDs by \citet{Decarli16} are
shown in Figure \ref{fig:cmgd} without correcting for the difference of
the CO-to-H$_2$ conversion factor used.
Since they used the constant conversion factor of 
$3.6~M_{\odot}~\rm(K~km~s^{-1}~pc^{2})^{-1}$, we also use the same
factor to directly compare with their results.
Resulting values are
$ 6.5 \times 10^7~M_\odot~{\rm Mpc}^{-3}$ (\citetalias{Tomczak}) and   
$ 5.1 \times 10^7~M_\odot~{\rm Mpc}^{-3}$ (\citetalias{Mortlock}) which are in the range of 
$ (3.4 - 12.3) \times 10^7~M_\odot~{\rm Mpc}^{-3}$ by \citet{Decarli16}.
It should be note that the CO luminosity corresponding to our $M_{\rm min}$ is close to the lowest CO luminosity used by \citet{Decarli16}.

The CMGD at $z \sim 0$ obtained here is about two times smaller than that by \citet{Obreschkow09a} ($(1.3 \pm 0.5) \times 10^7~M_\odot~{\rm Mpc}^{-3}$),
which is also shown in Figure \ref{fig:cmgd}. 
They derived the CMGD using the CO luminosity function by \citet{Keres03} and adopting the the CO luminosity and $B$-band luminosity dependent CO-to-$\rm H_2$ conversion factor.
\citet{Obreschkow09a} noted that \citet{Keres03} overestimated the CMGD due to the constant CO-to-${\rm H}_2$ conversion factor.
Although the CO luminosity corresponding to our $M_{\rm min}$ is close to the lowest CO luminosity used by \citet{Obreschkow09a},
the CMGD does not agree with that by \citet{Obreschkow09a}.
The cause for this discrepancy is not clear,
but it may be worth noting that if we use $f_{\rm mol}$ only by \citet{Boselli},
the CMGD agrees with that by \citet{Obreschkow09a}

We also show the semi-analytic model predictions by \citet{Lagos11} and \citet{Popping14} of the CMGD in Figure \ref{fig:cmgd}.
The model predictions roughly agree with the observational results.
However, the increase factor of the gas density from $z\sim0$ is rather smaller than that in this study.

In Figure \ref{fig:cmgd}, the best-fit CSFRD evolution by \citet{Madau} is also plotted,
which has an uncertainty of about $0.2$ dex as seen in their Figure 9.  
The CSFRD increases about ten times from $z\sim 0$ to $z\sim 1.5$ (contribution from ULIRGs at $z = 1 -1.5$ is not large ($10 -40$\%) (\citealt{Casey,Goto}).
Our results show the CMGD is also about ten times larger at $z\sim 1.4$ than at $z\sim 0$.
Thus the large CSFRD at $z =1 -2$ is considered to be due to the large CMGD at the redshift.
This would be reasonable if we recall that the SFR is mostly proportional to he molecular gas mass in a star-forming galaxy (e.g. \citealt{Sargent}).
This also implies that the star formation efficiency is similar at the both epochs on average.
\citet{Seko} pointed out that the star formation efficiency ($=$ SFR / molecular gas mass, or gas depletion time) is slightly larger at the higher stellar mass than at the lower mass in contrast to the trend seen in the local universe \citep{Saintonge11b,Boselli}.     
Although the trend is slightly different, the values are similar each other at the stellar mass of $\sim 2 \times 10^{10}~M_{\odot}$,
which is almost middle of the stellar mass sampled in this study.

In this paper, in order to constrain the CMGD at $z =1-1.5$ and that in the local universe,
we combined the average molecular gas mass fraction against the stellar mass and the stellar mass function of  star-forming galaxies at the redshifts.
By integrating down to $0.03 M^\ast$, the CMGD is derived.
The obtained CMGD at $z=1-1.5$ is $(6.8-8.8)~\times~10^7~M_\odot~{\rm Mpc}^{-3}$.
Although these values still have various uncertainties,
this CMGD at $z = 1- 1.5$ is about ten times larger than that in the local universe ($6.7~\times~10^6~M_\odot~{\rm Mpc}^{-3}$),
implying that the large CSFRD at $z=1-1.5$ is due to the large CMGD.
The CMGD at the redshift obtained in this study agrees with that recently obtained from integration of CO luminosity function,
indicating that the approach employed here is effective.\\\par

We would like to thank the referee for useful comments and suggestions.
K.O. is supported by Grant-in-Aid for Scientific Research (C) (16K05294) from Japan Society of the Promotion of Science (JSPS).
A.S. is supported by Research Fellowship for Young Scientists from JSPS.\\



\end{document}